\documentclass[sigconf]{acmart}

\AtBeginDocument{%
  \providecommand\BibTeX{{%
    \normalfont B\kern-0.5em{\scshape i\kern-0.25em b}\kern-0.8em\TeX}}}

\copyrightyear{2021}
\acmYear{2021}
\setcopyright{licensedcagov}\acmConference[SACMAT '21]{Proceedings of the 26th ACM Symposium on Access Control Models and Technologies}{June 16–18, 2021}{Virtual Event, Spain}
\acmBooktitle{Proceedings of the 26th ACM Symposium on Access Control Models and Technologies (SACMAT '21), June 16–18, 2021, Virtual Event, Spain}
\acmPrice{15.00}
\acmDOI{10.1145/3450569.3463573}
\acmISBN{978-1-4503-8365-3/21/06}

\usepackage{tabularx}
\usepackage{multirow}
\usepackage{array}
\usepackage{verbatim}
\usepackage{hyperref}
\usepackage{url}
\usepackage{tablefootnote}
\usepackage{subcaption}
\usepackage{balance}
\usepackage[font={small}]{caption}
\begin{document}
\fancyhead{}
\title{Analyzing the Usefulness of the DARPA OpTC \\Dataset in Cyber Threat Detection Research}

\author{Md. Monowar Anjum, Shahrear Iqbal}
\authornote{Corresponding Author: shahrear.iqbal@nrc-cnrc.gc.ca}
\affiliation{%
  \institution{National Research Council}
  \city{Fredericton}
  \state{New Brunswick}
  \country{Canada}
}
\email{{mdmonowar.anjum,shahrear.iqbal}@nrc-cnrc.gc.ca}

\author{Benoit Hamelin}
\affiliation{%
  \institution{Tutte Institute for Mathematics and Computing}
  \city{Ottawa}
  \state{Ontario}
  \country{Canada}}
\email{benoit.hamelin@cyber.gc.ca}








\begin{abstract}
Maintaining security and privacy in real-world enterprise networks is becoming more and more challenging. Cyber actors are increasingly employing previously unreported and state-of-the-art techniques to break into corporate networks. To develop novel and effective methods to thwart these sophisticated cyber attacks, we need datasets that reflect real-world enterprise scenarios to a high degree of accuracy. However, precious few such datasets are publicly available. Researchers still predominantly use the decade-old KDD datasets, however, studies showed that these datasets do not adequately reflect modern attacks like Advanced Persistent Threats (APT). In this work, we analyze the usefulness of the recently introduced DARPA Operationally Transparent Cyber (OpTC) dataset in this regard. We describe the content of the dataset in detail and present a qualitative analysis. We show that the OpTC dataset is an excellent candidate for advanced cyber threat detection research while also highlighting its limitations. Additionally, we propose several research directions where this dataset can be useful.
\end{abstract}





\maketitle

\section{Introduction}
Network intrusion and threat detection is an active research area in the cybersecurity domain. Researchers often rely on datasets that describe instances of cyber attacks. However, as enterprise systems grow in complexity and cyber attackers employ sophisticated techniques to breach into systems, it becomes increasingly important for publicly available datasets to reflect this picture. For instance, the recent Solarwinds attack on major corporations and government infrastructure across the United States \cite{oxfordaudacity} can be classified as an advanced persistent threat (APT) and currently very few publicly available datasets contain instances of such threats. There are also additional difficulties in constructing datasets that contain a diverse portfolio of attack scenarios. For example, privacy is one of the key concerns. It is often desired that datasets that are released publicly should be de-identified extensively to prevent accidental leakage of critical information. Strict requirements like this and other technical complexities often prevent organizations from releasing information rich cybersecurity datasets. 

Defence Advanced Research Projects Agency (DARPA) introduced the Transparent Computing (TC) program to develop technologies in order to gain high-fidelity visibility into the nature of today's computing environments. One of the primary missions of this program is to gain an understanding of APT attacks and develop experimental tools to counter these threats. In order to accomplish that goal, DARPA TC program constructed multiple datasets that contain large amount of malicious activities representing APT scenarios in an enterprise network environment \cite{darpaTC}. The latest iteration of these datasets is the Operationally Transparent Cyber (OpTC) dataset \cite{darpaOpTC}. This dataset contains more than 17 billion events from an enterprise network, describing both benign and malicious behaviors, through both network and host-level log telemetry. The sheer volume and richness of information present in the dataset makes it very useful to train traditional and deep learning models for detecting APTs or anomalies. 

Despite the dataset's enormous size, there is little to no explanation provided about the events in the dataset. This lack of documentation makes it difficult for cybersecurity researchers to properly analyze and identify important features for cyber threat detection. In this paper, we present a comprehensive documentation of the OpTC dataset. Our documentation includes an in-depth analysis of the network and host level events. We perform a data quality analysis by comparing the dataset with similar datasets. We also characterize malicious events and analyze the presence of class imbalance in the dataset. Finally, we describe possible future research directions that can take advantage of this dataset.

\paragraph{\textbf{Related Public Datasets}}
There are a number of similar cyber attack datasets that are being referenced in the contemporary literature. They vary in their nature as some of them represent data that are captured from the operational environment while some of them represent pseudo-real world events to model threats. There are also synthetic datasets which represent specific scenarios \cite{glasser2013bridging}. The relevance of a dataset is heavily time-dependant, as the attacks and tactics captured by comparatively older data often fail to mirror modern ones.

One of the datasets that is very similar to OpTC is the Los Alamos National Laboratory (LANL) Unified Host and Network dataset. It contains network and host activities of Los Alamos National Laboratory over the course of 90 days \cite{turcotte2017unified}. However, there is no specific malicious activity reported in this dataset. Previously in 2015, LANL released another cybersecurity event dataset that contains data from its enterprise network over 58 days. The dataset describes malicious activities performed by a red team, as noted in \cite{kent2015comprehensive}. Both of these datasets are referenced in cyber threat detection research.

The DARPA transparent computing program released several other datasets to the public \cite{darpaTC}. However, lack of understanding of the datasets due to insufficient documentation is one of the core reasons behind the datasets being not adopted widely in academic research. Meanwhile, older, arguably obsolete datasets with good documentation are still being used \cite{singla2020preparing}. For instance, the KDD99 dataset which is more than 20 years old is still used in intrusion detection research thanks to its comprehensive documentation. This is unfortunate as the attacks captured in the dataset bear little resemblance to modern threats \cite{ozgur2016review}.

The remainder of this paper is organized as follows. Section~\ref{sec:dscontent} describes the experiment testbed and actors involved in the construction of the dataset. We describe the content of the dataset and provide some statistics in Section~\ref{sec:dsdesc}. Section~\ref{sec:dsquality} contains a data quality assessment. Finally, Section~\ref{sec:dsrd} describes possible future research directions and we conclude in Section~\ref{sec:conclusion}.

\section{Experiment and Data Collection}
\label{sec:dscontent}
\subsection{Dataset Origins}
The DARPA-OpTC dataset is collected under a technology transition pilot study funded by Boston Fusion Corp.'s cyber APT scenarios for Enterprise Systems (CASES) project. The project's primary objective was to determine the scalability of the DARPA TC program without losing performance. The experiment consisted of one thousand hosts. Boston Fusion along with two other organizations from the TC program worked together to conduct the experiments and collected the data. Five Directions provided endpoint telemetry and BAE systems provided analysis over the data. Provatec played the role of both red team and test co-ordinator. This dataset represents only a subset of their activities.

\subsection{Experimental Setup}
The experiment testbed consisted of 1000 hosts with Windows 10 operating system. Each host of the experimental setup was equipped with a sensor. This sensor monitored the events and packed them into JSON records, which were dispatched to a central Kafka queue. Once received, the events were forwarded to an ETL server which aggregated and transformed them into the eCAR format. 

\subsection{The eCAR Data Model}
eCAR stands for Extended Cyber Analytics Repository. It is developed by Five Directions and is based on MITRE's CAR model \cite{mitreCar} that can describe an action over a host in a network. CAR is an event based model. Each event has three core components namely, \textbf{object}, \textbf{action} and \textbf{fields}. An object is an entity that has visibility from a network or a host. It can be subjected to actions performed by other entities in the host. Files or processes are examples of objects. A file can be read, written, created or deleted while a process can be created or terminated. Fields contain contextual information of a specific object/action pair. For instance, an event with the object "process" and action "create" has fields like command line, name, and process id. When combined together, an event \textbf{(object, action, fields)} can be perceived as a co-ordinate which represents a specific point in the temporal event space.

The eCAR format extends this model by creating a more complete representation of the events. For instance, the eCAR model contains \textbf{principal string} and \textbf{actor id} fields that describe which entity is performing the action on the current object. There are also timestamps and other fields that facilitate exploratory data analysis of events.

\section{Data Description}
\label{sec:dsdesc}
In this section, we perform a descriptive analysis of the dataset. First, we give an overview of the dataset content. Second, we quantify the distribution of events in the dataset. Last, we describe different objects and relevant fields of the objects. 

\subsection{Dataset Content}
In it's current format, The OpTC dataset contains around 1,100 gigabytes of data in the compressed JSON format. The JSON files contain eCAR events. The dataset is divided into 3 folders. 
\begin{enumerate}
    \item \textbf{ecar}:  This folder contains the eCAR events. There are three subfolders within this folder. They are named as \textbf{benign}, \textbf{evaluation} and \textbf{short}. The \textbf{benign} folder stores the normal activity captured between 19th and 23rd September. The \textbf{evaluation} folder stores events captured during the red team activity period, between September 23 and September 25. The \textbf{short} folder contains events that were captured during the exercise period but is missing values.
    
    \item \textbf{bro}: This folder contains data from Bro (now Zeek) sensors. 
    
    \item \textbf{ecar-bro}: This folder contains eCAR-formatted events annotated with bro-ids to link between records in the \textbf{ecar} stream and \textbf{bro} tables. 
    
\end{enumerate} 

The malicious activities performed during the evaluation period are described in the \textbf{OpTCRedTeamGroundTruth.pdf} file. The file contains details which can be used to label the malicious events in the evaluation folders.


\begin{table}[t]
    \footnotesize
    \centering
    \caption{Objects and their corresponding actions in the dataset. The percentage of each <Object, Action> pair is listed in the percentage column. The cells with the value 0.0 represent a rounded up value.}
    \begin{tabular}{|m{2cm}|m{3cm}|m{1cm}|m{1cm}|}
        \hline
        \textbf{Object} & \textbf{Actions} & \textbf{\%} & \textbf{Total \%}\\
        \hline
        \multirow{6}{2cm}{FILE} & CREATE & 1.1 & \multirow{6}{1cm}{12.4}\\\cline{2-3}
         & DELETE & 0.4 &\\ \cline{2-3}
         & MODIFY & 4.4 &\\\cline{2-3}
         & READ & 3.2 &\\\cline{2-3}
         & RENAME & 0.4 & \\\cline{2-3}
         & WRITE & 2.9 &\\
        \hline
        \multirow{3}{2cm}{FLOW} & MESSAGE & 21.6 & \multirow{3}{1cm}{71.7}\\\cline{2-3}
        & OPEN & 0.2 & \\\cline{2-3}
        & START & 49.9 &\\
        \hline
        HOST & START & 0.0 & 0.0\\
        \hline
        MODULE & LOAD & 3.9 & 3.9\\
        \hline
        \multirow{3}{2cm}{PROCESS} & CREATE & 0.1 & \multirow{3}{1cm}{8.6}\\\cline{2-3}
        & OPEN & 8.4 &\\\cline{2-3}
        & TERMINATE & 0.1 &\\
        \hline
        \multirow{3}{2cm}{REGISTRY} & ADD & 0.1 & \multirow{3}{1cm}{0.3}\\\cline{2-3}
        & EDIT & 0.2 & \\\cline{2-3}
        & REMOVE & 0.0 & \\
        \hline
        SERVICE & CREATE & 0.0 & 0.0\\ 
        \hline
        SHELL & COMMAND & 0.0 & 0.0 \\
        \hline
        \multirow{4}{2cm}{TASK} & CREATE & 0.0 &\multirow{4}{1cm}{0.0}\\ \cline{2-3}
        & DELETE & 0.0 &\\ \cline{2-3}
        & MODIFY & 0.0 &\\ \cline{2-3}
        & START & 0.0 & \\
        \hline
        \multirow{3}{2cm}{THREAD} & CREATE & 1.2 & \multirow{3}{1cm}{3.0}\\\cline{2-3}
        & REMOTE\_CREATE & 0.3 &\\\cline{2-3}
        & TERMINATE & 1.5 &\\
        \hline 
        \multirow{7}{2cm}{USER\_SESSION} & GRANT & 0.0 &\multirow{7}{1cm}{0.0}\\\cline{2-3}
        & INTERACTIVE & 0.0 &\\\cline{2-3}
        & LOGIN & 0.0 & \\\cline{2-3}
        & LOGOUT & 0.0 &\\\cline{2-3}
        & RDP & 0.0 & \\\cline{2-3}
        & REMOTE & 0.0 &\\\cline{2-3}
        & UNLOCK & 0.0 &\\
        \hline
    \end{tabular}
    \label{tab:obj_act}
\end{table}

\begin{table}
    \footnotesize
    \centering
    \caption{The categorization of the fields of objects in the dataset. The fields whose value persist across time is marked as \textit{persistent} and the rest are marked as \textit{volatile}.}
    \begin{tabular}{|m{1.9cm}|m{2.6cm}|m{1.5cm}|m{1cm}|}
        \hline
        \textbf{Object} & \textbf{Field} & \textbf{Permanent} & \textbf{Volatile}\\
        \hline
        \multirow{10}{1.9cm}{FLOW} & \verb!start_time! &  & \checkmark \\\cline{2-4}
         & \verb!end_time! &  & \checkmark \\\cline{2-4}
         & \verb!size! &  & \checkmark \\\cline{2-4}
         & \verb!src_ip! &  \checkmark & \\\cline{2-4}
         & \verb!dest_ip! &  \checkmark &\\\cline{2-4}
         & \verb!src_port! &  \checkmark & \\\cline{2-4}
         & \verb!dest_port! &  \checkmark & \\\cline{2-4}
         & \verb!l4protocol! &  \checkmark & \\\cline{2-4}
         & \verb!direction! &  \checkmark & \\\cline{2-4}
         & \verb!image_path! &  \checkmark & \\
        \hline
        \multirow{5}{1.9cm}{PROCESS} & \verb!command_line! &  \checkmark & \\\cline{2-4}
        & \verb!image_path! &  \checkmark & \\\cline{2-4}
        & \verb!parent_image_path! &  \checkmark & \\\cline{2-4}
        & \verb!user! &  \checkmark & \\\cline{2-4}
        & \verb!sid! &  \checkmark & \\
        \hline
        \multirow{5}{1.9cm}{FILES} & \verb!size! & & \checkmark \\\cline{2-4}
        & \verb!image_path! &  \checkmark & \\\cline{2-4}
        & \verb!info_class! &  \checkmark & \\\cline{2-4}
        & \verb!file_path! &  \checkmark & \\\cline{2-4}
        & \verb!new_path! &  \checkmark & \\
        \hline
        \multirow{3}{1.9cm}{MODULE} & \verb!base_address! & & \checkmark \\\cline{2-4}
        & \verb!image_path! &  \checkmark & \\\cline{2-4}
        & \verb!module_path! &  \checkmark & \\\cline{2-4}
        \hline
        \multirow{3}{1.9cm}{THREAD} & \verb!image_path! & \checkmark & \\\cline{2-4}
        & \verb!stack_limit! & & \checkmark \\\cline{2-4}
        & \verb!stack_base! & & \checkmark \\\cline{2-4}
        & \verb!source_pid! & & \checkmark \\\cline{2-4}
        & \verb!source_tid! & & \checkmark \\\cline{2-4}
        & \verb!target_pid! & & \checkmark \\\cline{2-4}
        & \verb!target_tid! & & \checkmark \\\cline{2-4}
        & \verb!subprocess_tag! & & \checkmark \\
        \hline
        \multirow{5}{1.9cm}{REGISTRY} & \verb!image_path! & \checkmark & \\\cline{2-4}
        & \verb!key! & \checkmark & \\\cline{2-4}
        & \verb!type! & \checkmark & \\\cline{2-4}
        & \verb!data! & & \checkmark \\\cline{2-4}
        & \verb!value! & & \checkmark \\
        \hline
        \multirow{5}{1.9cm}{TASK} & \verb!image_path! & \checkmark & \\\cline{2-4}
        & \verb!task_process_uuid! &\checkmark & \\\cline{2-4}
        & \verb!path! & \checkmark & \\\cline{2-4}
        & \verb!task_name! & \checkmark &\\\cline{2-4}
        & \verb!task_pid! & & \checkmark \\
        \hline
        \multirow{3}{1.9cm}{SHELL} & \verb!image_path! & \checkmark & \\\cline{2-4}
        & \verb!payload! & &\checkmark \\\cline{2-4}
        & \verb!context_info! & & \checkmark  \\
        \hline
        HOST & \verb!image_path! & \checkmark & \\
        \hline
        \multirow{4}{1.9cm}{SERVICE} & \verb!image_path! & \checkmark & \\\cline{2-4}
        & \verb!name! & \checkmark &\\\cline{2-4}
        & \verb!start_type! & \checkmark & \\\cline{2-4}
        & \verb!service_type! & \checkmark & \\
        \hline 
        \multirow{5}{1.9cm}{USER\_SESSION} & \verb!privileges! & \checkmark & \\\cline{2-4}
        & \verb!request_logon_id! &\checkmark & \\\cline{2-4}
        & \verb!request_domain! & \checkmark & \\\cline{2-4}
        & \verb!logon_id! & \checkmark &\\\cline{2-4}
        & \verb!requesting_user! & \checkmark &\\
        \hline
    \end{tabular}
    \label{tab:obj_feat}
\end{table}

\subsection{General Statistics of the Dataset}
The OpTC dataset has 17,433,324,390 events. These events split across 11 different object types, with 32 different (object/action) pairs. Every event has object specific fields that we categorize as either \emph{permanent} or \emph{volatile}. The fields for which the values persist across hosts during the whole experiment and are not ephemeral in nature are denoted as \emph{permanent}, while the other fields are denoted as \emph{volatile}. For instance, a process can have two different process IDs in two different hosts: we consider these IDs to be volatile. As an opposite, consider the command that has been executed to start a process. Regardless of the host, the process name (.exe) remains the same. Therefore, we categorize the process name as permanent. We believe that this categorization will help researchers identify features that yield more effective models.

In Table \ref{tab:obj_act}, we show that most of the dataset events consist of the FLOW, FILE and PROCESS objects. In total, these three events constitute more than 90\% of the entire dataset and the vast of majority of the information regarding the network dynamics are concentrated in these. As a result, it is important to have knowledge about the interactions between these three events and others. 

\subsection{Objects in the Dataset}
In this subsection, we will describe the dataset objects. In Table \ref{tab:obj_feat}, we show the fields of different objects and their categorizations (\emph{permanent/volatile}).

\subsubsection{FLOW} This object represents the occurrence of a communication between two hosts on the network, traditionally embodied in network flow records. It has three actions, namely MESSAGE, OPEN and START. Apart from the common fields, the FLOW object has a few additional fields such as start time, end time, size, source IP, destination IP, port, protocol specification, and image path. Image path identifies the path to the program that initiated that particular flow event. For the most part, the protocol specification field (\verb!l4protocol!) takes two values of 6 and 17 (TCP and UDP).

There is a significant difference between FLOW-START events and FLOW-OPEN/FLOW-MESSAGE events. The former are stand-ins in the eCAR stream to link to network flow records captured by the Bro sensor, through identifiers associated to objects in the \textbf{eCAR-Bro} directory. The other two designate phenomena that are actually occurring on hosts. The most frequent action type, MESSAGE, relate to communication through IP-based protocols. The OPEN action type is more opaque, and our analysis so far ties it to interprocess communications on a single host.

\subsubsection{FILE} This object represents file related activities in the hosts. It has
CREATE, DELETE, MODIFY, READ, RENAME and WRITE actions. This object has some FILE specific fields such as \verb!image_path!, \verb!info_class!, \verb!new_path! and \verb!file_path!. They contain host level metadata for file operations.

\subsubsection{PROCESS}This object represents host process events. The fields \verb!pid! and \verb!ppid! are complemented by \verb!tid!, which stands for thread id. These fields are host specific and volatile. It has a field \verb!sid! which encodes the user or account information that describes the level of privilege nominally associated to the process. It also contains a field named \verb!command_line! that provides context on how the process was started. 

\subsubsection{THREAD}Threads are usually created by the processes. The \verb!pid! and \verb!ppid! fields specify which parent process is responsible for the thread. These events can have three different actions: CREATE, REMOTE\_CREATE and TERMINATE. 

\subsubsection{MODULE}This object represents module load events. We researched extensively to figure out what it represents and found multiple references to Windows .NET Event Tracing Mechanism (ETW). It seems the event is created when a system library (dll) is loaded. Further investigation is needed to establish the effect of this object and it's actions.

\subsubsection{REGISTRY}This object represents a very small portion of the dataset (0.3\%). It has three actions: ADD, EDIT, and REMOVE. Some fields of this event are \verb!data!, \verb!value!, \verb!key! and \verb!type!. The first two are highly context specific and hence categorized as \textit{volatile} while the later two are context independent and \textit{permanent}.

\subsubsection{OTHER OBJECTS}Other than the above mentioned objects, the dataset contains USER\_SESSION, TASK , SHELL and SERVICE objects. Together they constitute less than 1\% of the dataset. The feature categorization of the object fields are provided in Table \ref{tab:obj_feat}.

\section{Data Quality Assessment}
\label{sec:dsquality}
In this section, we analyze the quality of the dataset. First, we compare the OpTC dataset with similar datasets available publicly. Then, we characterize the malicious events. Next, we discuss the class imbalance problem and ways this imbalance might affect the training of machine learning models. Lastly, we briefly mention the documented errors in the dataset. 

\subsection{Comparison with Similar Datasets} OpTC is a dataset which contains only network and host-level event logs. It does not have network packet-level information (e.g., packet captures). Therefore, we only compare with datasets that contain similar event logs. 

We compare the OpTC dataset with two datasets released by Los Alamos National Laboratory (LANL) in 2015 and 2018. Table \ref{tab:comp_net_flow} shows this comparison.\\

\begin{table}[h]
    \centering
    \footnotesize
    \caption{Comparison with similar datasets}
    \begin{tabular}{|m{2.7cm}|m{1.4cm}|m{1.4cm}|m{1.4cm}|}
    \hline
    Topic     & OpTC 2020 \cite{darpaOpTC} & LANL 2015 \cite{kent2015comprehensive} & LANL 2018 \cite{turcotte2017unified} \\
    \hline
    Number of events & 17,433,324,390 & 1,648,275,307 & 5,546,990,084  \\
    \hline
    Number of Malicious Events & 292367 (0.0016\%) & 749 (0.000045\%)& N/A  \\
    \hline
    Data Type & Network flow and host logs & Authentication records, net flows, process lifecycles, DNS requests & Network flow and host logs  \\
    \hline
    Timeline (Days) & 6 & 58 & 90 \\
    \hline
    Number of Hosts & 1000 & 17684 & $\approx$17500 \tablefootnote{Derived from a graph in \cite{turcotte2017unified}}  \\
    \hline
    \end{tabular}
    \label{tab:comp_net_flow}
\end{table}

\begin{figure*}[t!]
    \begin{subfigure}[t]{0.33\textwidth}
        \centering
        \includegraphics[height=1.5in]{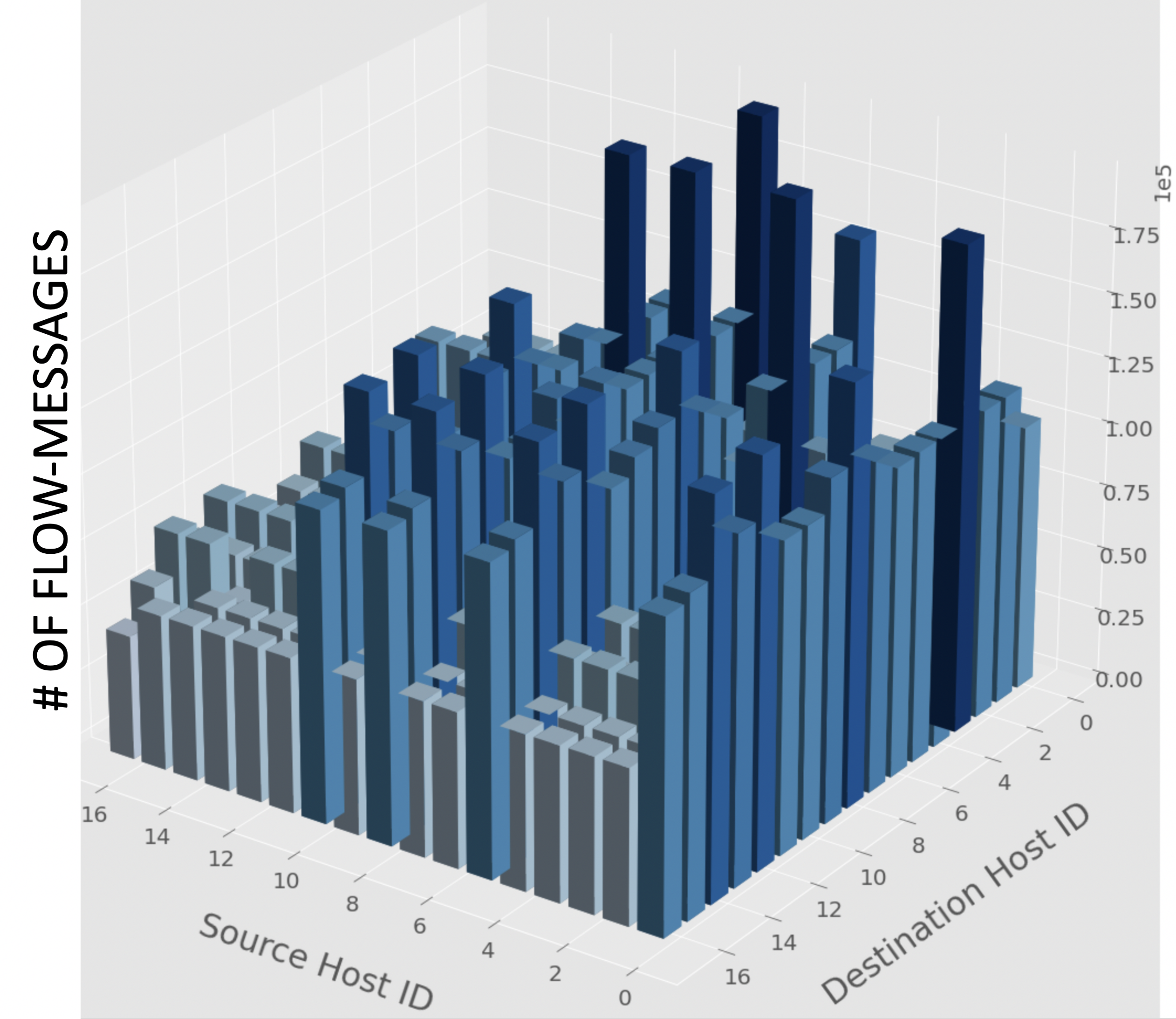}
        \caption{Day 1}
    \end{subfigure}
    \begin{subfigure}[t]{0.33\textwidth}
        \centering
        \includegraphics[height=1.5in]{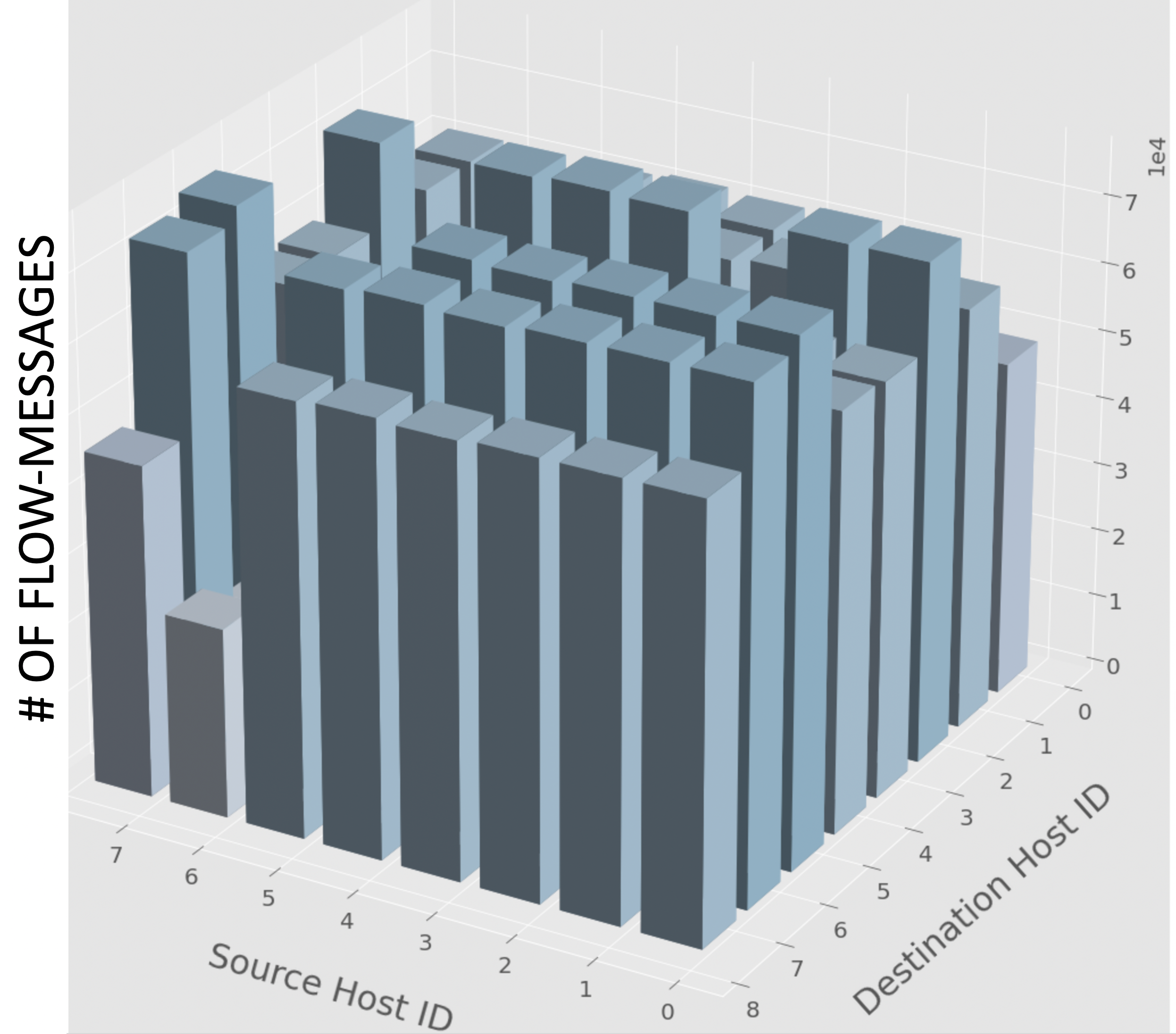}
        \caption{Day 2}
    \end{subfigure}
    \begin{subfigure}[t]{0.33\textwidth}
        \centering
        \includegraphics[height=1.5in]{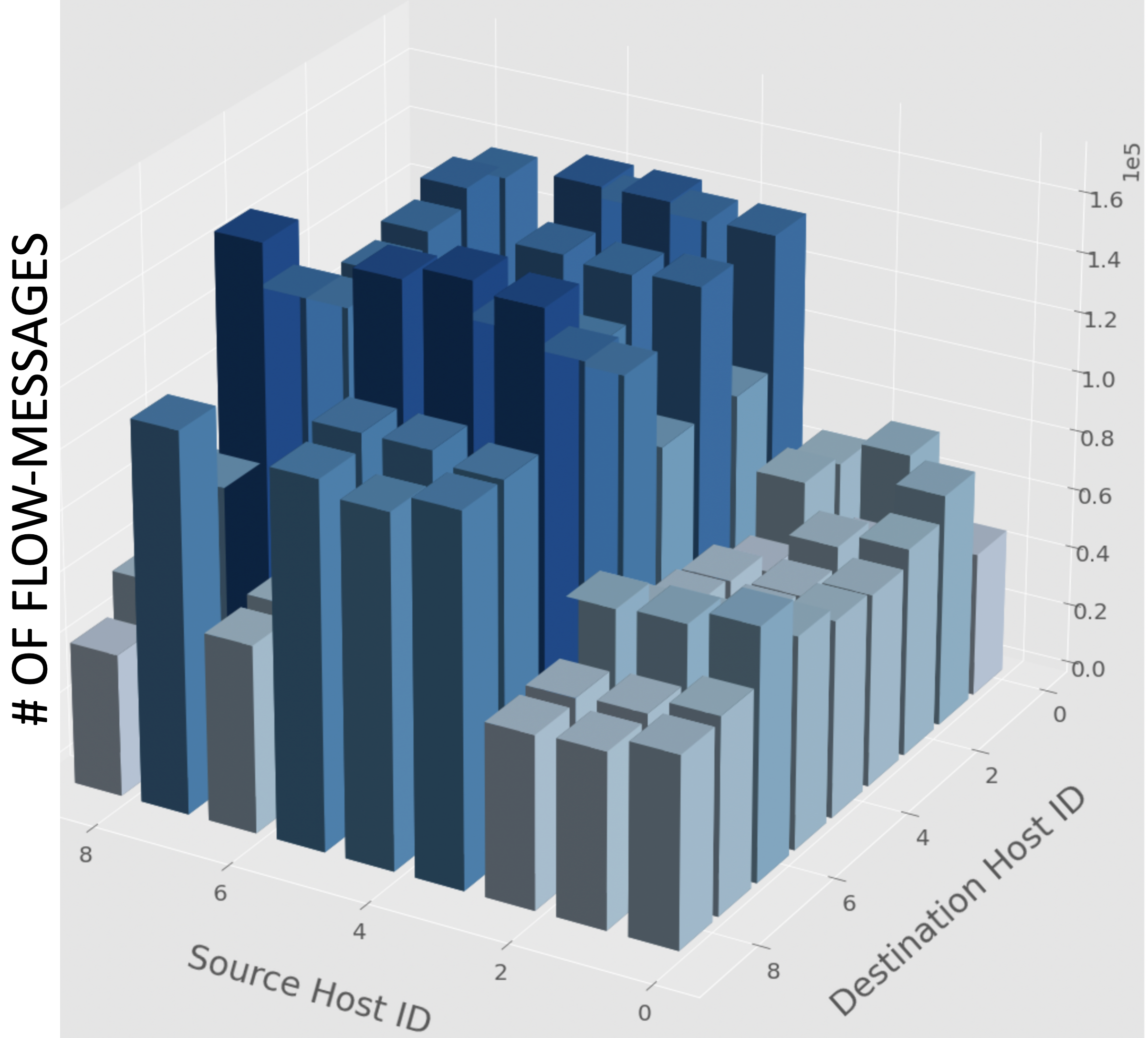}
        \caption{Day 3}
    \end{subfigure}
    \footnotesize
    \caption{Network FLOW-MESSAGE distribution between malicious hosts during the evaluation period. Day 1 represents the powershell empire staging scenario, Day 2 represents the custom powershell empire scenario and Day 3 represents malicious software update scenario.}
    \label{fig:flow_distribution}
\end{figure*}

The cyber activities in the OpTC dataset encompass a broad range of network flow and host activities. The host activities include file, process, thread, and module operations. This variety and richness of information, coupled with the higher density of events describing attacker actions, enable comprehensive investigation into both benign and malicious activities which are not possible in other datasets. For instance, the LANL 2018 dataset does not have any event that shows the shell commands executed by any process on the host machine. The LANL 2015 dataset discarded most of the contextual information regarding processes, in particular their filiation relationships. In contrast, the OpTC eCAR data format contains significant amount of contextual information that facilitates fine-grained feature engineering.

\subsection{Characterization of Malicious Events} The OpTC dataset contains malicious events performed over the course of three days during the evaluation period. The first day portrays a powershell empire staging scenario. It contains examples of initial foothold establishment, lateral movements, and privilege escalations. The second day contains events that include data exfiltration via Netcat and RDP. The third day contains instances of malicious software upgrades. All of these malicious activities are consistent with the behaviour of an advanced persistent threat \cite{juels2012sherlock}. It makes this dataset an excellent source for advanced persistent threat detection or APT stage classification research. 

This particular aspect gives this dataset a distinct advantage over contemporary cybersecurity datasets. For instance, the LANL 2015 dataset only tags login events with nominal red team labeling \cite{kent2015comprehensive}. Extending these labels to further events is complicated and ultimately heuristic, resulting in a poor basis to assess the performance of attack detection methods. The LANL 2018 dataset does not have any documented red team activities which severely compromises its utility in advanced persistent threat detection, limiting its applicability to the development of baseline models.

In order to demonstrate the presence and movements of attackers within the network, in Figure \ref{fig:flow_distribution}, we provide visualizations of the network activities among compromised hosts during the evaluation period. Day 1 contains the most examples of events related to APT. It also involves the maximum number of hosts (3.4\%). Day 2 and Day 3 involve 1.6\% and 1.8\% hosts respectively. This distribution of network flows among the malicious hosts demonstrates how advanced persistent threats establish their foothold in one of the hosts, perform lateral movements to escalate their privilege within the network and eventually compromise the target host to exfiltrate data and achieve other attack goals.

\begin{figure}
    \centering
    \begin{subfigure}[ht]{0.25\textwidth}
        \centering
        \includegraphics[height=1.5in]{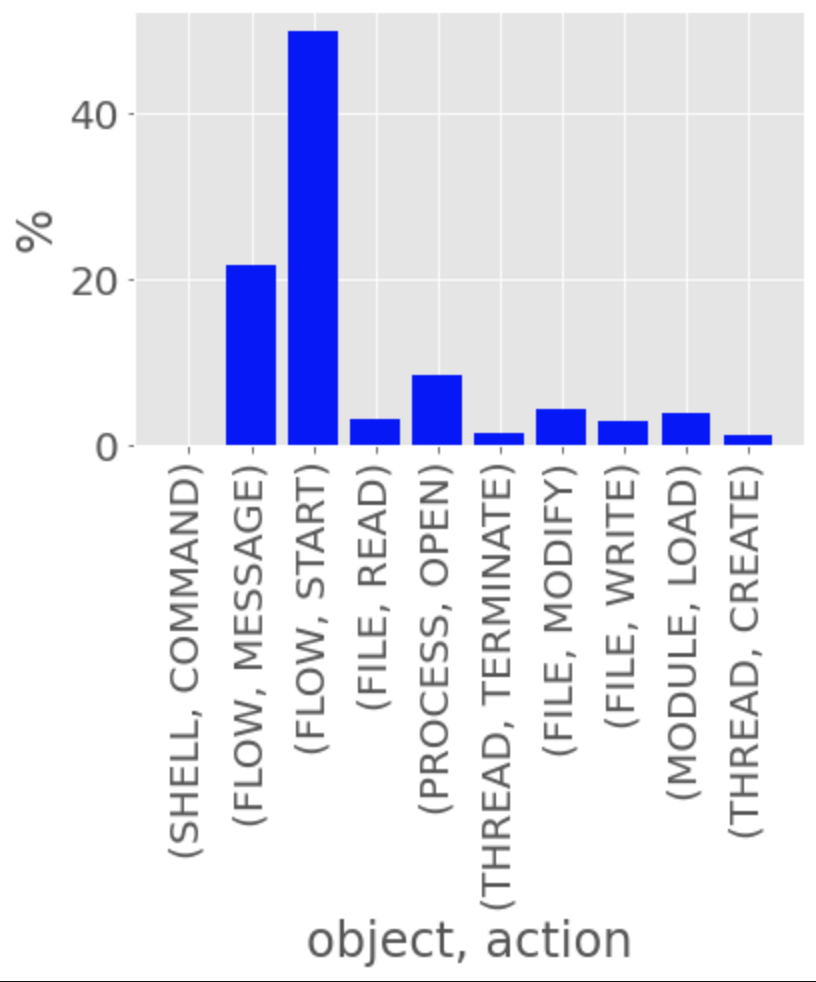}
        \caption{Benign Events}
    \end{subfigure}%
    \begin{subfigure}[ht]{0.25\textwidth}
        \centering
        \includegraphics[height=1.5in]{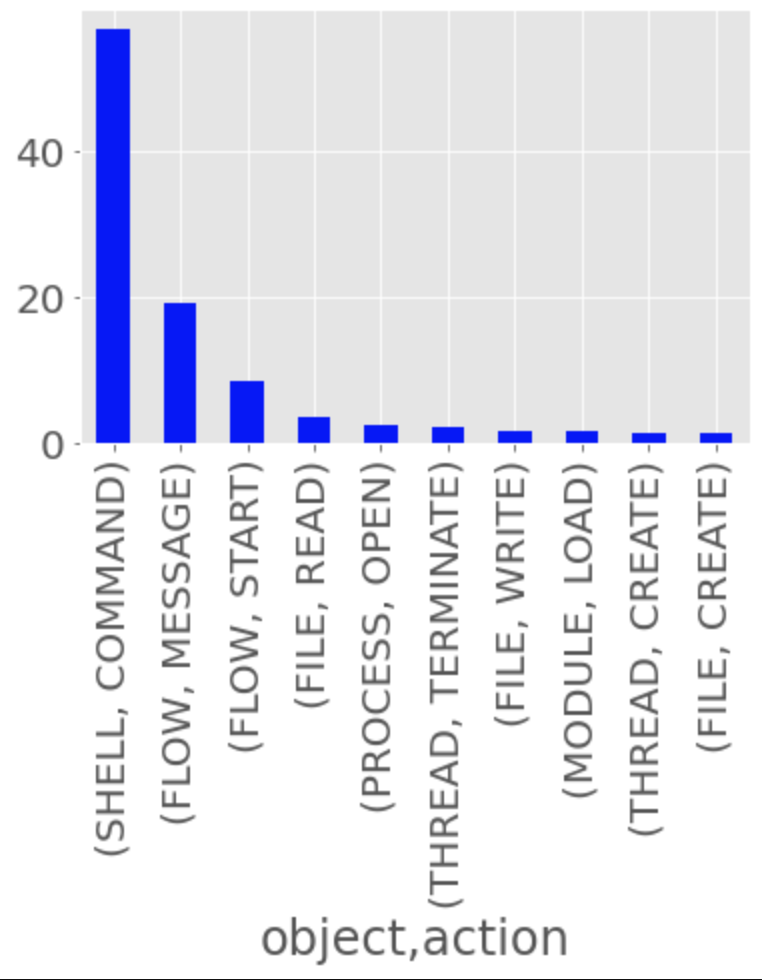}
        \caption{Malicious Events}
    \end{subfigure}%
    \caption{Distribution of Benign and Malicious events}
    \label{fig:event_distribution}
\end{figure}

In Figure \ref{fig:event_distribution}, we show the difference between the distribution of benign events and malicious events in the OpTC dataset. We see that malicious events have greater percentage of the SHELL-COMMAND event while in the case of benign events, they are practically non-existent. Taking a closer look at the ground truth data revealed that most of the red team activities involved \textit{windows management instrumentation}, \textit{mimikatz} and commands like \verb!lsadump! and \verb!ipconfig!. Basically, the malicious agents ran a lot of shell commands in power shell which resulted in the high percentage of SHELL-COMMANDs.

\subsection{Class Imbalance} The dataset contains more than 17 billion events in total. However, the number of malicious events is slightly less than 0.3 million (approximately 0.0016\%). This imbalance in number between benign and malicious events presents a significant challenge in anomaly and threat detection. Specifically, it becomes difficult to train deep learning models on the imbalanced dataset since the performance of deep learning models in terms of generalization tend to degrade with imbalanced class representations \cite{buda2018systematic, japkowicz2000class}. However, this imbalance is prevalent in operational enterprise networks.

In enterprise networks, advanced persistent threats usually infiltrate a system long before the actual malicious activities even take place. These threats have attack models that leave minimal footprint and they frequently use zero-day exploits to perform malicious activities \cite{alshamrani2019survey, han2020unicorn}. Therefore, the class imbalance in this dataset is consistent with the real-world scenario.

The class imbalance problem in datasets has been investigated in the literature by some researchers. Wheelus \textit{et al.} \cite{wheelus2018tackling} used bagging, undersampling and synthetic minority oversampling on the UNSW-NB15 dataset \cite{moustafa2015unsw} to tackle the problem. Their experiments on this network-flow-only dataset have shown significant performance improvement in terms of classification accuracy. Presumably, similar techniques can be applied on the OpTC dataset when developing machine learning models, taking into account the heterogeneous nature of the data stream.

\subsection{Errors in the Dataset} According to our understanding, there are a few issues with PROCESS objects and their sources. The sources were not de-conflicted properly in some cases which may result in an erroneous entry in the \verb!actor_id! field. Another error is the \verb!acuity_level! of the FLOW object which has the value of 0. For every other object, this value is within the range of 1 to 5. 

\section{Research Directions}
\label{sec:dsrd}
Cybersecurity threat detection and attack stage classification are active research areas and the OpTC dataset can be utilized to gain insight into these problems. In this section, we discuss several future directions where researchers can use the OpTC dataset.
\balance
\subsection{Anomaly Detection}
Anomaly detection in a cybersecurity dataset can be defined as the problem of detecting patterns that do not fit the distribution of the events present in the dataset \cite{ahmed2016survey, akoglu2015graph}. This problem is distinct from outlier detection since network and host events are often related in an implicit way, which is not always captured in vector embeddings. A promising area of research stems from the tendency of using deep learning models to implicitly learn the most useful features for the task they are designed to accomplish. The training of such deep networks requires very large amounts of data, which made prior datasets unsuitable. We hypothesize that the much larger data volume of the OpTC dataset will better support this work.

Another problem with the previous anomaly detection models was high false-positive rates \cite{turcotte2017unified}. We believe that the richness of the feature set of the OpTC dataset can help in this regard. While both the LANL and OpTC datasets contain event logs, the latter sports a larger number of event-specific features than the former. The high level of detail facilitates the interpretation of sequences of activity identified by detection models. They allow precise characterization of normal phenomena initially flagged as anomalies, enabling the iterative removal of large swathes of events that are part of baseline activities. Normal data thus accounted for and shaved off of a model's input reduce the class imbalance during training and increase the size of the intersection between the set of identifiable anomalies and that of actual malicious activities. This stands to improve systematically the specificity one may expect from detection methods.

\subsection{Representation Learning}
Representation learning of a network event graph in the context of cybersecurity is an emerging research area \cite{usman2019survey}. Contributions so far have focused on provenance graphs \cite{han2018provenance}, as well as approaches based on neural networks. Provenance-based approaches provide a robust representation of network events that offer insights into the collective network behavior, at the cost of some detection sensitivity compared to NN approaches. These, however, are practically black box, hampering the explanation of the representation in cyber defense terms. While the OpTC dataset provides reliable ground truth for good representation learning, its wealth of attributes also offers paths towards improved explainability. 

\subsection{Process Tree Building} An important area of research enabled by the OpTC dataset involves \emph{process and event trees}. Enterprise operating systems are process-oriented in nature. Every single process can be mapped to another process that initiated it; every state change can be mapped to a process having caused it to occur. Therefore, there exist hierarchical relationships between processes and events, encoded in eCAR attributes \textbf{actorID} and \textbf{objectID}, which can be made explicit to analyze data phenomena. Irregularities in these trees can be marked as anomalies and can be further scrutinized as originating from malicious cybersecurity events. Much like in the analysis of the data as a chronological sequence, the very low density of events describing malicious activities and the long tail of the distribution of events raise significant challenges to the specificity of malicious activity detection. We presume to address these challenges similarly as we suggested above.

Finally, process and event trees encode sequential information regarding events that are chronologically distant but semantically close. This makes them a valuable representation for APT detection algorithms as advanced persistent threats often mirror this trait. In fact, APT processes trigger sporadically to avoid raising any red flag. Therefore, building process trees can be a step in the right direction to formulate efficient representation of events for detecting APTs.

\section{Conclusion}
\label{sec:conclusion}
Cybersecurity datasets that represent a real-world scenario are very important in supporting cyber defense research. However, good datasets with proper documentation are scarce. In this work, we documented the recently introduced DARPA OpTC dataset. We described the dataset content and performed a data quality analysis. Our work showed that the dataset contains a large volume of cyber events that represent the real world to a high degree of accuracy. The intrusion experiment captured therein emulate advanced persistent threat tactics, which is highly valuable in cyber defense research. We have also discussed its usefulness in anomaly detection and representation learning for detecting advanced persistent threats. We hope that our work will encourage potential cybersecurity researchers to perform extensive analysis on the dataset to develop tools for combatting advanced cyber threats. 
\begin{acks}
We acknowledge the help of Mike Van Opstal from Five Directions and Mr. Tejas Patel from DARPA CASES project for providing us with additional technical details on the construction and semantics of the dataset. 
\end{acks}

\bibliographystyle{ACM-Reference-Format}

\bibliography{SACMAT2021}

\end{document}